# Cyber Autonomy: Automating the Hacker
## – Self-healing, self-adaptive, automatic cyber defense systems and their impact to the industry, society and national security

Ryan K. L. Ko, UQ Cyber Security, University of Queensland
St Lucia, Queensland, Australia
ryan.ko@uq.edu.au

In 2016, the Defense Advanced Research Projects Agency (DARPA) hosted the Cyber Grand Challenge, a competition which invited participating finalist teams to develop automated cyber defense systems that can self-discover, prove, and correct software vulnerabilities at real-time. As the competition progressed, the systems were not only able to auto-detect and correct their software, but also able to attack other systems (other participants' machines) in the network. Even though the competition did not catch much mainstream media attention, the Cyber Grand Challenge proved the feasibility of cyber autonomy, stretched the imagination of the national and cyber security industries and created a mix of perceptions ranging from hope to fear. Since 2017, IT vendors started moving towards 'security automation' (as evidenced by the recent rise in automation vendors at the RSA conferences (2017, 2018) - the world's largest cyber security vendor trade show at San Francisco) – the first leap towards cyber autonomy. We also witnessed the increased effort towards automating active cyber defense tools, including situation awareness, deception, penetration testing and vulnerability assessment tools. Like other industries disrupted by automation, these trends have several implications on national security and the private industry. The trend created a major disruption to the currently in-demand penetration testing professional sector – changing manual penetration testing into a sunset industry overnight. While penetration testers still fetch relatively high salaries at the time of writing, the emergence of several cyber artificial intelligence (cyber-AI) techniques are starting to change business models and national security human resource planning: from needing to hire large teams of penetration testers and security assessment consultants (from junior to experienced staff) into a new business model which utilises so-called cyber-AI and automation tools managed by just a handful of extremely-skilled professionals. This chapter sets the context for the urgency for cyber autonomy, and the current gaps of the cyber security industry. A novel framework proposing four phases of maturity for full cyber autonomy will be discussed. The chapter also reviews new and emerging cyber security automation techniques and tools, and discusses their impact on society, the perceived cyber security skills gap/shortage and national security. We will also be discussing the delicate balance between national security, human rights and ethics, and the potential demise of the manual penetration testing industry in the face of automation.

## 1   Introduction

In 2016, the Defense Advanced Research Projects Agency (DARPA) hosted the Cyber Grand Challenge (Song & Alves-Foss, 2015), a competition which invited participating finalist teams to develop automated cyber defense systems that can self-discover, prove, and correct software vulnerabilities at real-time – without human intervention. For the first time, the world witnessed hackers being automated at scale, i.e. cyber autonomy (Brumley, 2018). As the competition progressed, the systems were not only able to auto-detect and correct their software, but also able to attack other systems (other participants' machines) in the network.

Even though the competition did not catch much mainstream media attention, the DARPA Cyber Grand Challenge proved the feasibility of cyber autonomy, stretched the imagination of the national and cyber security industries and created a mix of perceptions ranging from hope to fear – the hope of increasingly secure computing systems at scale, and the fear of current jobs such as penetration testing being automated.

Since 2017, IT vendors started moving towards 'security automation' (as evidenced by the recent rise in automation vendors at the RSA conferences (2017, 2018, 2019) (RSA Conference, 2019)– the world's largest cyber security vendor trade show at San Francisco) – the first leap towards cyber autonomy. We also witnessed the increased number of automated active cyber defense tools, including deception, penetration testing and vulnerability assessment tools. Like other industries disrupted by automation, these trends have several implications on national security and the private industry.

Many vendors are touting network automation programs with existing security information and event management (SIEM) tools as cyber autonomy. Others would label security 'playbooks' – hardcoded heuristics which script responses according to triggers or a combination of triggers – as automation. An example of a 'playbook' would be a pre-programmed workflow of actions responding to a variety of cyber-attacks (e.g. a response to a denial of service attack or a network port scan by an unknown source).

These examples are still a distance from the true potential and vision for cyber autonomy. The holy grail for cyber autonomy is that we can deter attacks and patch vulnerable computing systems at real-time, at scale **and** without disruption to normal operation. The crux of this is the assumption that a computing system handles abstract and virtual executions, and hence has fewer physical limitations and boundaries for dynamic remediation.

However, from a practical implementation viewpoint, this assumption does not hold strong ground. Software systems, particularly those running critical infrastructure, emergency services, and 24/7 manufacturing, have very complex dependencies, and do not have the luxury to be turned off and patched during downtime due to their operational demands. For example, the software running a nuclear power plant should not be shut down abruptly.

The dilemma between the need to patch system vulnerabilities and the need to maintain business or operational continuity also places pressure on software migration processes and time. Software migration (or modernization) is the current practice of modernizing software to a newer version. The interdependency of processes and software makes this a challenging change management process. Proponents of cyber autonomy would argue that with cyber autonomy, the need for systems (in particular, critical infrastructure systems) to be modernised would be reduced since the self-healing aspects of cyber autonomy will address vulnerabilities without disrupting business-as-usual.

In the author's opinion, it is important to note that cyber autonomy is differentiated against the concept of 'hack-back' (Messerschmidt, 2013), which is linked to defense organisations proposing an autonomous system intentionally hacking back an attributed source of cyber-attack with the aim of decisively limiting or terminating future attacks. Cyber autonomy is more likened to the analogy proposed by Barton Miller *(a.k.a. the 'father' of fuzzing (an automated software testing technique))*, who likened dynamic binary instrumentation (Bernat & Miller, 2011) (a field within cyber autonomy) to that of "fixing a car while we are driving it".

This chapter is organised into eight sections. Following the introduction in Section 1, Section 2 covers the cyber autonomy challenge by first discussing the scale and temporal challenges of cyber defense and the consequential need for cyber autonomy. Following this, in Section 3, we propose a novel framework of maturity levels required to achieve full cyber autonomy. Section 4 surveys current cyber defense automation approaches and Section 5 discusses the impact of automated cyber offense. Section 6 covers the ethical and human rights challenges posed by full cyber autonomy and we discuss human resource trends and provide recommendations in Section 7, before concluding the chapter in Section 8.

## 2   The crux of the problem

Why is cyber autonomy a turning point in the history of computing systems? One would be able to understand the implications better by first understanding the current 'fire-fighting', unsustainable nature of cyber (and particularly, software) security.

Whenever a software is created and released, one can assume two factors occurring across most cases: (1) that the software will have some unknown bug or vulnerability despite best efforts in testing due to the abstract nature of software development, and (2) that the software engineers working on the code have little knowledge or training in skills required to write code securely.

Until recent years, most programming language coding training at private training institutes or universities have seldom focused on writing code in a way which prevents weaknesses and logical errors which lead to cyber security vulnerabilities. In fact, several of the problems such as buffer overflow (Crispan, et al., 1998) have been around for decades with no end in sight. Nevertheless, software is usually released before they are fully ready – in order to meet investment release deadlines or business targets.

These factors have contributed to software released to their users in a best-effort way. Even if the software has been formally verified for potential errors, up to the requirements of international standards such as the Common Criteria (Herrmann, 2002), the risks are only minimised at best. The recent Boeing 737 Max fatalities (Johnston & Harris, 2019) resulting from software bugs in the cockpit which automatically and repeatedly forced the aircraft to nosedive are testament to that. In most cases, we are unable to be absolutely certain that there will not be new vulnerabilities discovered in the future. Contrast this to the building and construction industry standards and practices which assures reliability, safety and accountability of the building industry. On this note, legal penalties for insecure software engineering practices have been mooted, but that is a separate discussion which will be discussed in Section 5.

## 2.1 Reactive software vulnerability remediation and its challenges

Due to widespread insecure software engineering, the discovery of software vulnerabilities (commonly called 'bugs') is a common occurrence and some bugs offer entry points for cyber criminals to enter software and access or breach sensitive data. Software released (ranging from operating systems, e.g. Windows, Linux, MacOS, to mobile phone apps) would have the following typical stages of vulnerability remediation (See Figure 1):

- X: Time from start of vulnerability discovery to the first public announcement/responsible disclosures
- Y: Time for vulnerability patch development
- Z: Time for patching the vulnerable software in deployed systems

Within the software industry, the time period X in Figure 1 would typically start from the discovery of vulnerabilities and conclude at the first public announcement by security professionals through disclosures or the appropriate software vendors. The vulnerabilities declared are usually included and categorised into MITRE's Common Vulnerability Enumeration (CVE) list (MITRE, 2005) and eventually into the National Institute of Standards and Technology (NIST) National Vulnerability Database (NVD) to articulate the severity of the vulnerability (Booth, et al., 2013).

**Current situation: X+Y+Z > A or B**

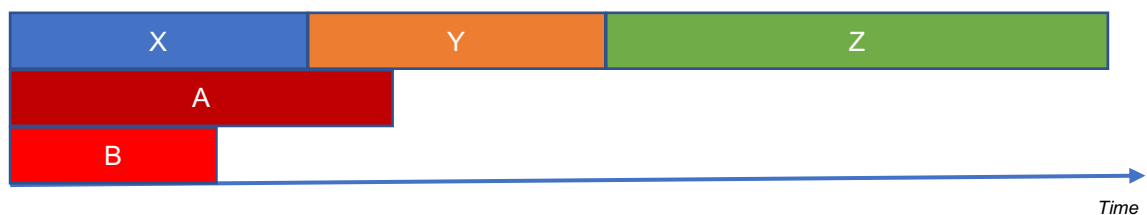

Legend:
- X: Time from vulnerability discovery to public announcement/responsible disclosures
- Y: Time for patch development
- Z: Time for patching the vulnerable software
- A: Time for cyber criminals, based upon knowledge of publicly-disclosed vulnerabilities, to develop exploits to take over (i.e. pwn-ing) a system
- B: Time for cyber criminals, based upon knowledge of zero-day vulnerabilities to take over (pwn-ing) a system

Disaster: (X+Y+Z) > A

*Figure 1: Current timeframes for software vulnerability discovery and remediation*

The CVE list and the NVD serve as public service announcements which enable security experts to understand the vulnerability and develop patches to remedy the problem or patch the vulnerability (i.e. time period Y in Figure 1). CVEs and the NVD are also resources for several cyber security tools and global cyber incident/emergency response teams.

Organisations and individuals would then update or patch their systems (i.e. Z in Figure 1). Unfortunately, sometimes Z can range from a few days to even months or years – depending on the complexity, processes and dependencies of the systems. In 2018, statistics (Rapid7, 2019) show that it takes an average of 39 days to patch a vulnerable system (i.e. time period Z). It was reported that medium severity vulnerabilities took an average of 39 days to patch while interestingly, low severity vulnerabilities took an average of 54 days. The oldest unpatched CVE (common vulnerability enumeration) took 340 days to remediate.

Even when the vulnerabilities are declared and X is short, like a double-edged sword, the CVE and NVD information may also serve as intelligence sources for cyber criminals to develop new ways and software scripts known as exploits to enter vulnerable systems which are not (yet) patched for the new known vulnerabilities. The time from knowing the new CVE entry to the development of an 'exploit' (i.e. a way to automate the takedown of a machine via the known vulnerability) to the subsequent success cyber-attack is captured as 'A' in Figure 1.

Compared to Z, the exact statistics for the length of X and Y are usually hard to measure. If a vulnerability was discovered by cyber criminals and are not declared to the vendors, this would be known as a zero-day vulnerability. These cyber criminals may either choose to develop their own exploits or choose to sell their newly developed exploits (e.g. average price of bug bounties and exploits at around $3400 (Lemos, 2019)) or vulnerability knowledge on anonymous platforms such as the Dark Web. It must be noted that even government state agencies purchase such knowledge. The time from the discovery of the vulnerability to the development of exploit would be captured in Figure 1 as 'B'. With reference to Figure 1, a zero-day vulnerability would also mean that the time period X is as long as it can be, until an individual or organisation disclose the vulnerability publicly, or to the affected software vendor. Usually, B is shorter than X.

As shown in Figure 1, the current situation is that a typical organisation would take (X+Y+Z) to patch their known vulnerabilities. The resiliency of an organisation is increased when it is able to shorten Y and Z significantly. It would be disastrous if **A<(X+Y+Z)** or **B<(X+Y+Z)** for an organisation. In fact, these unfortunate scenarios are still happening today, as evidenced by the WannaCry ransomware attack (Mattei, 2017), where unpatched hospital equipment in NHS hospitals were unable to function after the attack.

The time taken to remediate the vulnerability also opens a window of opportunity for cyber criminals to take over systems. In some situations, Y and Z – both within the control of organisations – can take an extremely protracted time. An organisation with a long Z for its production systems needs to work on the effectiveness, knowledge and efficiency of its IT personnel and management. Therefore, our goal is to make X+Y+Z as short as possible (see Figure 2). This is why cyber autonomy – with its automated vulnerability detection and remediation (i.e. self-healing) – is such an attractive option.

**Ideal Situation: (X+Y+Z) < A or B**

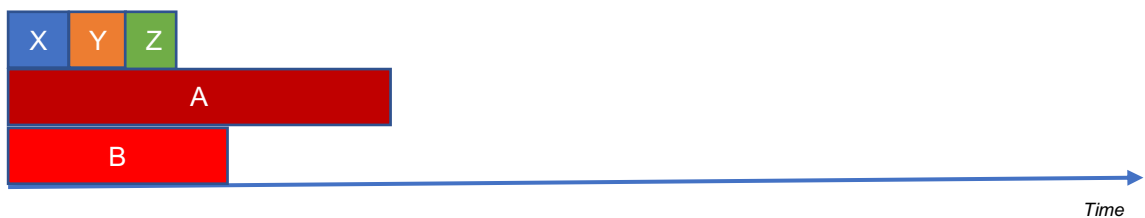

*Figure 2: Ideal situation in software vulnerability discovery and remediation*

## 2.2   Scale of the threat landscape

We have considered the situation from the temporal perspective. Let us now look at the situation from a scale perspective at a global context.

### 2.2.1   Rate of vulnerabilities discovered versus rate of software produced – mobile, web applications

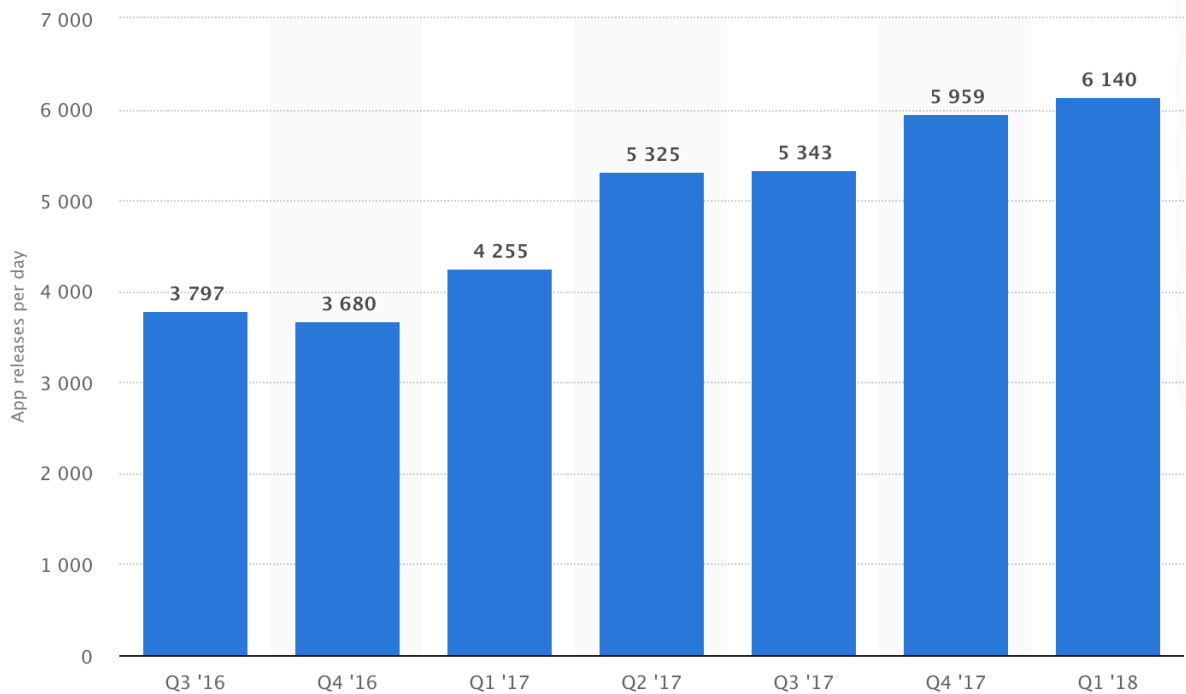

*Figure 3: Average number of new Android app releases per day (Q3 2016 to Q1 2018) (Clement, 2019)*

The statistics in Figure 3 show the average number of new Android app releases per day from the third quarter 2016 to the first quarter 2018. From these figures, it can be seen that during the last measured period, an average of 6140 mobile apps were released through the Google Play Store each day. This is equal to about 8.5 new apps released on Google's Android each minute. Such growth patterns are not expected to stop.

In contrast to the rate of new apps developed, the rate of discovery of vulnerabilities are lagging behind significantly. This builds a strong case for automation, especially because the software vulnerability identification and remediation processes can be done at rate faster than production. When the rate of such security automation increases at a higher rate than that of software produced, we will reach an equilibrium point, followed by the reduction of vulnerabilities exposed to potential criminals.

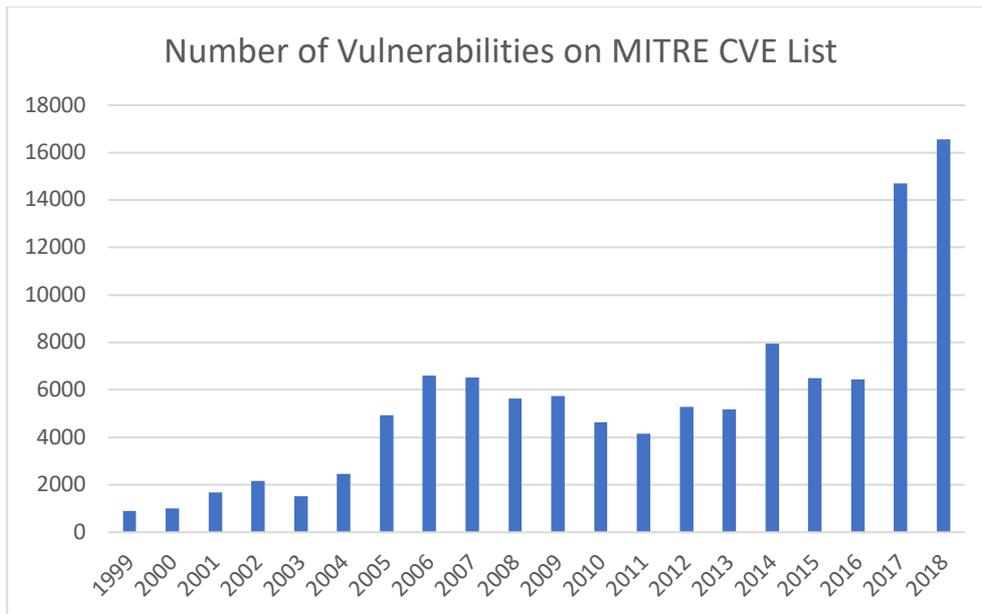

*Figure 4: Number of vulnerabilities reported on MITRE CVE List. Src:(CVE Details, 2019)*

The best global indicators of vulnerabilities would be the number of vulnerabilities obtaining their relevant CVE numbers on the CVE list. As shown in Figure 4, since the advent of the CVE in 1999, the number of reported vulnerabilities to the CVE list has increased almost exponentially. It is important to point out that the increase could actually be even higher than this, as these records do not include zero-day vulnerabilities, which are used "in the wild", without the knowledge of manufacturers or users. Solutions for software security would need to be sustainable. Problems generated at a machine rate need to be matched with solutions formulated at a machine rate.

### 2.2.2    Rate of malware discovered

The AV-TEST Institute registers over 350,000 new malware and potentially unwanted applications (PUA) every day (AV-TEST - The Independent IT-Security Institute, 2019). These new malware or unwanted applications are then examined and classified according to their characteristics and saved. Visualisation programs then transform the results into diagrams that can be updated and produce current malware statistics. This translates to about 4 new malware or PUA registered per second.

### 2.2.3    Comparing malware discovery rate to speed of skills training and education

When we look at the statistics of the threats generated, it soon become obvious that the best strategy to overcome the volume and veracity of the new vulnerabilities discovered should not be a linear one. With no intention to sound cliché, there needs to be a strategy to 'change the game'. This could be achieved through new generations or types of weaponry to overcome adversaries or changing the battlefield into a 'new chess board' to change the rules of engagement. History in the physical realms of security have shown us glimpses of how these strategies work. From the introduction of aircrafts in wars, to the development and use of nuclear weapons in the second world war, new responses seem to be the only viable solution to an unsustainable future.

Within the Web security space, we can see a simple way to 'change the game' in responses to web spam visits by bots. CAPTCHA (Von Ahn, et al., 2003), a randomly generated picture which can only be interpreted by human users, was introduced to web forms to cut down the number of automated spam bots which were crawling websites and sending spam information into organisations. The introduction of CAPTCHA, a simple additional feature, stopped web spamming. In a similar fashion, two-factor authentication techniques, which require users entering an additional authentication factor (e.g. a passcode sent to the user's mobile phone) on top of the usual password authentication, has also rooted out several phishing campaigns.

While there is a strong push for the cyber security skills gap to be address, and the author agrees for the need to educate more cyber security professionals across all skill levels, there is only so much training and education can do to fill the gap which is growing at the rates shown above.

Logically and practically, it takes a much longer period of time for a person to be equipped with cyber security skills and experience relevant to manually finding vulnerabilities and remedying them. Typically, the time taken to approve the deployment of new patches into production systems, and the time taken to test new patches released, would not be able to catch up with the deluge of software and malware. The rate of threats and software released are at a much higher rate and pace, and there needs to be a shift from reactive to proactive defense. This is why active defense through full cyber autonomy is the only viable option we have so far.

## 3  The path to full cyber autonomy

So, what is *full* cyber autonomy? In its essence, this means a point or time when technical solutions match or overcome the scale and temporal challenges posed by malware and vulnerabilities described in the previous section. Full cyber autonomy is achieved when any computing user – regardless of his/her technical background – is able to help or protect himself/herself against cyber-attacks and attribute the attack sources. This is supported by automated tools allowing him/her to make judgement calls or plan cyber security strategies without the need to rely on technical experts or a need to actively and continuously observe visualizations for security events.

Achieving full cyber autonomy means the availability of tools which empower users to make strategies and implement them automatically. It is important to note that autonomy is not full automation, but more of a man-machine symbiosis (Licklider, 1960), which combines the strategic strengths of the human user with the scale and speed of software and automation solutions.

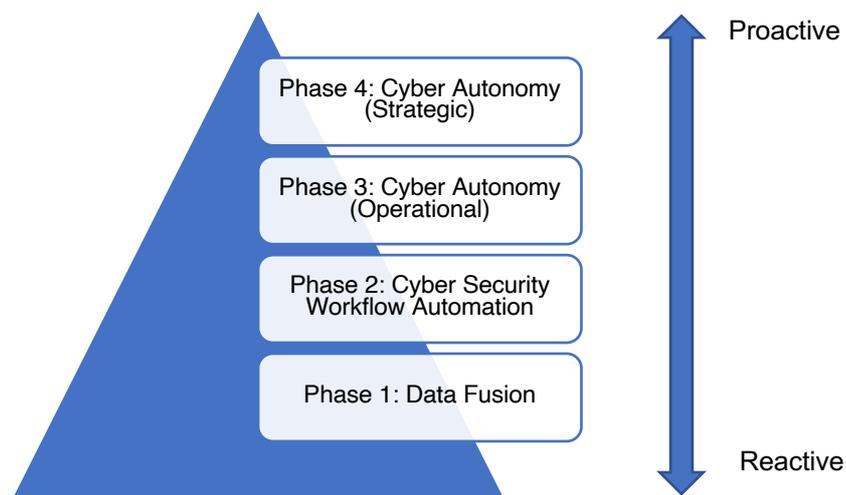

Figure 5: Cyber autonomy maturity phases

We propose four phases of maturity for full cyber autonomy:

- **Phase 1: Data Fusion** – At this maturity phase, tools are able to bring/fuse disparate data together to assist users with awareness of 'what goes on behind the scenes' within and across machines – within, outside and across the sovereign or organisational boundaries (e.g. cloud computing services on Amazon Web Services or Microsoft Azure cloud). Most of the current solutions are based on historical data and not real time data. Since the late nineties, the emergence of security information and event management (SIEM) tools have assisted system administrators with abilities to collate and correlate large volumes of log datasets from disparate devices, routers and computers across an organisation's infrastructure landscape. The SIEM tools were able to present the risk levels (typically color-coded to traffic light colors, with red denoting 'extremely serious' and green denoting healthy systems) and automate remediation responses (e.g. blocking traffic from malicious IP addresses). Most current cyber security solutions are in Phase 1 as they automatically integrate disparate data sources but are unable to provide the capabilities of the following phase.

- **Phase 2: Cyber security workflow automation** – This phase involves cyber security tools integrating disparate data sources and automatically responding to simplistic scenarios, using workflows which can be pre-programmed as rules based on Boolean logic like *"if this is true, then do x; otherwise, do y"*. Since the early 2010s, the application of machine learning into (extremely) large datasets, epitomised by Google DeepMind's application of deep learning techniques using neural networks to perform so-called predictions, to highly-complex but bound problem sets such as strategy games Go (Gibney, 2016), have inspired a new generation of so-called 'AI for Cyber Security' solutions. The above-mentioned DARPA Cyber Grand Challenge in 2016 played a significant role in showing the feasibility of the then-mostly academic techniques (e.g. static and dynamic analysis of software, symbolic execution, dynamic instrumentation, fuzzing) for real-world applications. At the time of writing, the cyber security products and services industry is now moving towards Phase 2 at a steady pace, and other some sectors such as defense and technology are at the cusp of moving into Phase 3 (at least an aspirational level).

- **Phase 3: Cyber autonomy (operation)** – At this phase, tools will augment intelligence into mostly-manual operational processes of identifying issues in infrastructures at real time – thereby reducing the need to hire staff to audit software for vulnerabilities manually. The key difference with Phase 2 is that Phase 3 includes context awareness, and the continuous awareness of organisational structure updates, processes, and policies. Phase 3 achieves the vision put forth in the eighties and nineties on 'continuous auditing and monitoring' for accountability of actions (Kogan, et al., 1999), and users are empowered with true automation of lower-level tasks. Since there is no explicit way to model business processes from the lenses of cyber security, the industry is still lacking description capabilities for automation, and is naturally some distance away from Phase 3. Current techniques are mostly automating responses designed through the coding of heuristics rather than a truly intelligent plan automatically generated from artificial intelligence planning *(a branch within artificial intelligence where systems take in goals in consideration of constraints and generate new plans based on the reasoning algorithms they run on)*. When Phase 3 is achieved, systems will be resilient to adversarial machine learning attacks (i.e. garbage-in, garbage-out problems faced by machine learning algorithms) and will be able to handle both syntactic (i.e. raw string of characters) and semantic (i.e. meaning derived from the string of characters) levels of abstraction.

- **Phase 4: Cyber autonomy (strategic)** – At this phase, the user will be able to articulate *strategies* in plain language, supported by artificial *general* intelligence applied into cyber security scenarios. It is at this phase that we can *truly* claim that the user is able to help themselves protect against or respond to cyber incidents – achieving *full* cyber autonomy. At Phase 4, the systems will be able to handle pragmatics, on top of semantics and syntactic levels of abstraction (Searle, et al., 1980). At Phase 4, the systems should also be able to handle ethical expectations and behavioural norms. Cyber norms (Finnemore, 2017) within and across jurisdictions would also be catered to.

As one looks at the state and future of cyber security tools through the lenses of the four maturity phases, it can be observed that the technologies display a shift from reactive security to proactive security as we progress from Phase 1 to Phase 4. As we progress from Phase 1 to Phase 4, we can also see an increasing degree of autonomy from the users' viewpoint. The main goal is to counter the current asymmetry where solutions are outweighed by the rate and variety of attacks with the aim of achieving parity between solutions and attack rates.

# 4  Automating cyber defense – current approaches

From the above maturity phases, it is clear that the current solutions fall into the Phase 2 (cyber security workflow automation) capabilities. Recently, some industry solutions labelled as 'artificial intelligence for cyber security' products are trailblazing into Phase 3 (cyber autonomy – operation). To have an appreciation of the opportunities and challenges created from the road to full cyber autonomy, we will now look at the current approaches of Phase 2. While automatic discovery and patching of vulnerabilities (Bernat & Miller, 2012) take centre stage in cyber autonomy, there are many more areas which complement the dynamic vulnerability discovery and patching process. These areas attempt to limit the capabilities of the hacker and outsmart the hackers through deception and a changing network topology.

## 4.1  Cyber Deception

In several high-profile advanced persistent threats (APT) such as Stuxnet (Langner, 2011) or the Bangladesh bank heist (Kaspersky Labs, 2018), cyber criminals tend to covertly lurk within the network infrastructure for a long period of time – ranging from months to years – before they strike. The main goal of cyber deception technologies is to trap and defeat stealthy cyber attackers through luring or deceiving them into attacking seemingly vulnerable (but fake) infrastructure. This builds on earlier related but more simplistic network security technologies known as honeypots or honeynets (i.e. multiple honeypots) – which collect network traffic of attackers on machines intentionally exposed to potential network attacks. Cyber deception technologies aim to collect information about the movements and attack strategies of an attacker, rather than simply stopping access. The goal is to understand their modus operandi well enough to devise strategies and solutions so that the attackers cannot come back in again using the same techniques.

When deployed well, and supported by a strong team of experts, deception technologies can potentially reduce the asymmetry between solutions and attacks and turn the tables on attackers. That said, just like honeypots, attackers will soon understand that they are entering a deception technology environment after some time, rendering the deception technologies outdated and increasing the pressure on deception technology vendors to come up with more and more sophisticated cyber defense strategies.

## 4.2  Automated penetration testing/vulnerability assessment

As the number of vulnerabilities discovered increase exponentially over time, there is an increasing trend where companies use automated tools to conduct vulnerability scanning on their servers and applications. Like a building inspector or an auditor looking through a checklist of potential issues, the automated penetration testing and vulnerability assessment tools will systematically scan all programs on all relevant computers for known vulnerabilities informed by the NVD list. The current tools are good at flagging the known vulnerabilities and presenting the potential problems to the users but are clumsy when it comes to the next crucial step: remediation of the vulnerabilities. The process is efficient but not very effective since the results of such scans are often fraught with false positives. This means that there is still a necessity to rely on technical experts at some point in the process. As the tools get continuously refined, there are several further implications to the training and the skill levels of the workforce and these will be discussed in Section 7.

## 4.3  Playbooks for security workflow orchestration

With solutions gaining momentum towards Phase 3 from 2017, solution providers and vendors are starting to offer add-on capabilities which allow security professionals to pre-program response workflows to initial attack symptoms such as denial of service, port scanning, unauthorised or suspicious user privilege escalation, and so on. The steps for countering threats are programmed into workflows as so-called playbooks. Playbooks are orchestrations of steps (think flowcharts) which will help to automate and save precious time so that security professionals can focus on strategic tasks. Since mid-2019, the concept of playbooks has increasingly been adopted within incident response communities and organisations. In the author's opinion, playbooks are a necessary precondition for Phase 3 and beyond. For a playbook to work effectively, there is a need for human oversight to describe the workflows within the playbooks. This predefinition of a context is likened to the field of AI planning, where before an algorithm can come up with solutions in the form of a 'plan', human operators need to first describe planning problem 'domain' (e.g. a crane moving containers in a port), the domain's 'atomic actions' (hold and release containers, up, down, left right) and their 'preconditions and effects' (Ghallab, et al., 2004).

## 4.4  Moving target defense (MTD)

To level the playing field between attackers and defenders, moving target defense (MTD) techniques (Jajodia, et al., 2011) will help cyber infrastructure landscape and settings (e.g. changing network topologies, evolving user account structures, etc.) to evolve. This will increase the level of effort needed by attackers to perform reconnaissance or lurk within the environments. The network topology learned by the criminal today will not be the same topology tomorrow, even though it is still the same target organisation. In essence, MTD's goal is to constantly change the characteristics of the infrastructure so that attackers would become ineffective or discouraged from attacking their targets. Currently, the concept is appealing but it is hard to implement at scale; it will be a few more years before this technology consolidates. That said, the concept of MTD promises to flip the asymmetric advantage towards the defenders.

# 5 Offensive cyber autonomy and its implications for national security and international regulations

As we progress towards full cyber autonomy, we begin to observe that the same capabilities used for autonomous *defensive* cyber security can also be used for *offensive* cyber security. The risk of weaponization of cyber automation also becomes more likely, and like other weapons, there needs to be clear legislation, regulations or guidelines around their usage and disposal. Weaponization of cyber-related knowledge and software already exists. In fact, to a certain degree, weaponization is increasing through the thriving markets within the anonymous Dark Web, including through the trading of zero-day vulnerabilities, 'hacking as a service' and cyber-attack software – with buyers ranging from nation states to mafia organisations (Radianti, et al., 2009).

To our best knowledge, while efforts have been made to apply existing international law (e.g. Law of Armed Conflict) to cyber weapons e.g. the Tallinn Manual 1 and 2.0, most current legal definitions and regulations of weapons do not cater to cyber weapons – a type of abstract weapon which is agnostic to geographical boundaries and jurisdictions. The definitions and laws (both domestic and international) are also unable to handle complex cyber-attacks which blur the lines between crime and acts of war. In the words of Liivoja, Naagel and Väljataga (Liivoja, et al., 2019), "regulatory complexity increases as legal concerns become compounded". A good example is the February 2016 Bangladesh bank heist where hackers (attributed to the so-called Lazarus group) attempted to illegally transfer close to US\$ 1 billion out of the Bangladesh central bank's account with the Federal Reserve Bank of New York via the SWIFT network. While the actual cyber-attack was a transnational crime scenario, the crime had further geopolitical implications since the 'earnings' made by the Lazarus group will eventually make it back to their paymaster, North Korea (Kong, et al., 2019). Eventually, such funds and earnings will likely fund the North Korean leaders' lifestyle, nuclear enrichment programs and ballistic missile tests.

The Bangladesh bank heist example also brings out the critical need for government agencies to be able to coordinate together without being limited by their usual mandates or scope. In several countries, the role of protecting national critical infrastructure (e.g. banks) falls into a grey area between the usual responsibilities of defence forces, the police, and the signals or intelligence agencies. In the aftermath of a cyber-attack launched through autonomous technologies, who should lead the response, and if an interagency task force was already set up, how is it empowered and how effective can it truly be? It is critical that governments work proactively to answer these questions rather than be overtaken by events; being proactive will enhance their abilities to rapidly respond to cyber-attacks and reduce security and economic costs.

Another issue for national security and law enforcement is the (in)ability to prove or disprove intent. Cyber-attacks, especially those automated to the sophistication of Phase 2 maturity levels and above, will likely combine attacking machines owned by the attackers and machines owned by victims who are not even aware that their machines were controlled by attackers. In the 2016 Mirai botnet attack (Antonakis, et al., 2017) which turned vulnerable unpatched devices into remotely controlled 'bots' as part of a large scale international botnet attack, several of the computers and devices launching the worldwide attacks are caused by ordinary unpatched devices ranging from medical devices to Internet-connected closed circuit television (CCTV) cameras. Several of the owners of these vulnerable devices are unaware that they contributed to the attacks and had no direct intention to launch or participate in the Mirai attack. Despite their ignorance or lack of intent, their devices nevertheless contributed to the inaccessibility of several prominent sites such as Twitter and Netflix. Given this situation, would the owner of the vulnerable device(s) be complicit and accountable for the cybercrime? If so, how is intent proven?

Another challenge for regulation of cyber autonomy is the combination of the transnational nature of cyber-attacks and the existing difficulty in attributing the source of cybercrime. The global regulations currently do not facilitate or help address this challenge. The Budapest Convention on Cybercrime (Council of Europe, 2004) does exist, but non-member nations do not have the benefit of exchange of information derived by the 67 signatories. The effectiveness of the Budapest Convention on Cybercrime also has not been empirically measured or studied. At the time of writing, with 195 countries in the world, non-member nations make up the bulk of the ratio which does not have access to the agreed cooperation and access to cybercrime information. They need to rely on other forms of communication for intelligence. While the usefulness of the Budapest Convention is recognised by most countries including non-member nations, many non-members view the requirement to change/align their respective national laws to align to Budapest Convention expectations as an obstacle for practical or timely implementation. For some countries, it may require much more effort to navigate through their political environments. As such, some alternative methods are the usage of their member nation status in the INTERPOL (190 member nations), or through one-to-one collaborative relationships with the countries they are seeking information from or with.

Cybercrime is increasingly automated and syndicated. The Blackhole Exploit Kit (Krebs, 2016) was a good example of criminals selling 'hacking as a service' and providing fancy dashboards for their subscribers displaying the computers infected by their malware. As cybercrime grows in complexity and volume, the basic flow of information for investigation and crime fighting purposes is also hampered by the difficulties of attributing the true source of cyber-attacks. For example, a distributed denial of service (DDoS) attack may be launched from computers at Country A, but these computers could be controlled by computers in Country B, or worse, controlled by Country C and so on. When provided with the evidence of an attack through digital forensics and network logs, an investigator usually needs to find further clues that may debunk the assertion that attacks were from the said source. When trying to prove the source, some laws also contain technical requirements which mean that some of the critical digital evidence may not be usable in courts of law. For example, ordinary logs from cloud computing servers have several duplicates across other cloud servers (a normal practice in large global clouds like Google or Amazon Web Service).  These are transient in nature and do not lay a strong foundation for data integrity and tamper prevention, especially when compared to the traditional digital evidence of physical hard drives found in someone's computer. Unless an organisation uses servers which require high integrity or 'chain of trust' from a cryptographic chip in the physical machines running the cloud instances, most logs collected do not satisfy strong evidentiary requirements.

Finally, because software is often released without being 100% assured, there is always a chance of the cyber automation tools (or weaponry) going out of control. Unintended consequences may occur and, even if the software is formally (i.e. mathematically) proven to a high level of assurance, there is still some chance of untested what-ifs. The recent example of the Boeing 737 Max cockpit software problems which caused several air crashes and fatalities (Pasztor, et al., 2019) was a good example of automation software going out of control in the form of unintended consequences. When we progress from Phase 1 to Phase 4 maturity levels, we need to develop policies and regulations which will address the liabilities and engineering standards imposed on software as an industry – and this goes beyond cyber automation software. In the case of a fully autonomous cyber software going rogue and taking down infrastructure because of a previously undiscovered bug, who or what can be held responsible and accountable? Making software engineers criminally responsible for badly engineered software has been a contentious and divisive topic, but there needs to be more analysis and discussion of this issue.

# 6   Implications for ethics and human rights

During the development of the technology and the progression from Phase 1 to Phase 4, adversarial nations may gain access to the automation through stealing intellectual property and knowledge.  This could result in an increasingly fast-paced battle/war between the automation haves and the have-nots. The chasm between the haves and have-nots will also widen, quite possibly forcing the situation where haves become the masters and the have-nots become subjects or slaves. For example, when every device you technically operate is under the control of a more powerful and Phase 4-ready organisation or country, actors are likely to lose control over their activities, rights and civil liberties. The recent shut-down of the Ukrainian power plants through cyber-attacks by an overseas state actor (Booz Allen Hamilton, 2019) demonstrated this possibility. The ethical boundaries and expectations around full cyber autonomy remains an open topic but the prospects do not appear to be good. This conclusion does not emerge from a pessimistic viewpoint but rather comes from a consideration of the practical realities. In the face of adversaries with unequal terms of engagement or ethical expectations, the incentive to maintain the same ethical ground will naturally diminish over time – especially after a few high-profile incidents which will create pressure for decision makers to act swiftly – often with compromises to privacy or ethics.

With more powerful and faster autonomous technologies for cyber defense and attack, there will naturally be more capabilities for individuals without technical expertise and organisations to scour the Web and other open source intelligence (OSINT) sources for reconnaissance purposes. In this context ensuring personal privacy is likely be an increasingly uphill battle. Much like the ethical and human rights expectations imposed on the usage of biomedical tools, such as gene editing, cloning or stem cell research, there will need to be sustained advocacy around ethics, human rights and norms of responsible behaviour for the engineering and usage of these cyber autonomy technologies. The state-of-the-art in cyber security research suggests that ethical approval processes in the most developed nations consider mostly biological ethical risks. The same attention is not given to ethics for cyber security, data privacy and data science technologies or experiments. A re-examination of ethics approval processes at the research and engineering stages could be a good baby step towards widely accepted global ethics and norms and the protection of civil liberties in autonomous systems for cyber defense.

# 7 Implications for human resource trends and recommendations

Full cyber autonomy will disrupt the global skills shortage problem and morph the skills gap problem from a focus on quantity to a focus on quality. With the current Phase 2 to Phase 3 transition, there is a stronger need for truly excellent penetration testers with actual bug bounty experience and tangible track records. The experience issue is a catch twenty-two situation, since it is impossible to train all fresh graduates to be the cream of the crop with a range of experience in bug and threat hunting. With the automation of vulnerability assessments and penetration testing tools, traditional penetration testing companies that depend on manual penetration testing face becoming a sunset industry.

The media plays an interesting and important role in shaping perceptions of job roles. When we observe the semiotics of current cyber security news articles, job brochures, or marketing material for tertiary qualifications, it is common to see images with locks and keys, handcuffs, a digital avatar wearing a hoodie jacket typing on a keyboard, dark images, and of course, backgrounds with 1's and 0's. The industry recently realised that this semiotics created a skewed demand for 'hacking' jobs, instead of an actual cyber security profession. A recent design competition by the Hewlett Foundation (Sugarman & Wickline, 2019) to encourage new cyber security marketing material towards non-hoodie images was a sign that the industry has started to consider a more accurate representation of the diversity of the roles and jobs in cyber security. The move towards the diversity of roles and stakeholders in cyber security should also consider semiotics which encourage people to consider jobs as controllers of automation tools, and a higher emphasis on strategic and analytical thinking. For all the advantages of full cyber autonomy, computers alone are insufficient to handle the increasingly complex cyber security attacks and challenges. We need skills and facilities, training, man-machine symbiosis and to leverage the strategic skills of a humans to monitor (and at times, prevent) mishaps from happening when automated tools perform unintended actions.

Generally, national skills policies aimed at increasing cyber hygiene and creating an awareness of career options are good, but they are not going to be sustainable for the future. There is currently too little emphasis on developing skills which allow professionals to constantly adapt to the new (and daily) challenges posed by cyber-attacks launched quickly by cyber autonomy tools. Skills such as research, data analysis, or the application of evidence-based techniques to solve complex issues are usually under-represented or under-championed. While industry stakeholders typically do not expect to hire cyber security job applicants with masters or PhD degrees, this could be a mistake from the macro viewpoint and in the longer term because they will miss out on the research thinking and mindsets of these cohorts. Cyber security skills and education policies must aim to develop a new generation of skills which are not only confined to penetration testing, they should encourage training researchers and toolmakers who can create automated cyber security tools.

Related to toolmaker training is the critical need for well-engineered cyber security software and tools. To put this into perspective, the buffer overflow problem commonly found in software is now decades old, but few schools actually teach coding with an emphasis to prevent such fundamental vulnerabilities. We need to advocate for a culture training new generations of prospective cyber security professionals with the ability to read and write secure code, backed by automation tools – supported by exposure to skillsets from other disciplines including but not limited to criminology, political science, and business management.

# 8  The road ahead

In thie chapter, we illustrated the asymmetric scale and speed of attacks (i.e. 4 new malware per second) versus the current mostly manual defensive techniques and time it takes to conduct skills training. This sets the context for the urgency for cyber autonomy and highlighted current gaps of the cyber security industry. To understand the current status in the context of *full* cyber autonomy, we proposed a novel framework with four phases of maturity for full cyber autonomy. It was observed that at the time of writing, the cyber security industry is currently at Phase 2 of our framework, while many organisations are moving towards Phase 3 (at least on an aspirational level). Phase 4 will be realised when technology achieves artificial general intelligence. We reviewed emerging cyber security automation techniques and tools, and their impact on society, the perceived cyber security skills gap/shortage and national security. We also highlighted the lack of global alignment using the Budapest Convention as an example, and discussed several challenges, including the difficulty in proving criminal intent since many owners of compromised devices do not even realise that their unpatched devices are contributing to global attacks.

Overall, the road towards full cyber autonomy has more positives than negatives, but only if we can emphasize creating a mindset to engineer cyber defense tools rather than attack tools. Autonomous cyber defense tools should also focus on reversing the asymmetry between the rate of attacks and efficiencies of defense. The ethical principles around designing autonomous cyber defense tools should be well-thought through and mirror industries such as the biomedical sector which regulates technologies such as cloning and gene editing.